\documentclass[aps,prl,twocolumn,groupeauthor,showpacs]{revtex4}
\usepackage{amsmath,amssymb,latexsym}
\usepackage{graphicx}
\graphicspath{{../fig/}}
\usepackage{txfonts,mathrsfs}
\usepackage{comment}
\DeclareSymbolFont{symbols}{OMS}{cmsy}{m}{n}
\DeclareSymbolFont{largesymbols}{OMX}{cmex}{m}{n}
\newcommand{\bm}[1]{\boldsymbol #1}

\newcommand{\KK}{{\bm K}}

\newcommand{\QQ}{{\bm Q}}

\newcommand{\CC}{\mathcal{C}}

\newcommand{\latt}{\mathcal{L}}
\newcommand{\lattC}{{\mathcal{L}'}}
\newcommand{\bzC}{{\mathcal{B}'}}
\newcommand{\bz}{{\mathcal{B}}}

\newcommand{\Tc}{\mathcal{T}_\CC}
\newcommand{\Tr}{\text{tr}}

 \newcommand{\pseudoG}{{\mathcal{G}}}

\newcommand{\pseudoS}{{\mathfrak{S}}}

\newcommand{\spinor}{\psi}
\newcommand{\sspinor}{\Psi}
\newcommand{\conv}{\ast}

\newcommand{\sd}{\downarrow}
\newcommand{\su}{\uparrow}
\newcommand{\tp}{{ t^\prime}}

\begin{document}

\title{Ultrafast Quenching of the Exchange Interaction in a Mott Insulator}

\author{J.~H.~Mentink}
\email{Johan.Mentink@mpsd.cfel.de}
\author{M.~Eckstein}

\affiliation{Max Planck Research Department for Structural Dynamics, University of Hamburg-CFEL, 22761 Hamburg, Germany}
\date{\today}

\begin{abstract}
We investigate how fast and how effective photocarrier excitation can modify the exchange interaction $J_\mathrm{ex}$ in the prototype Mott-Hubbard insulator. We demonstrate an ultrafast quenching of $J_\mathrm{ex}$ both by evaluating exchange integrals from a time-dependent response formalism, and by explicitly simulating laser-induced spin precession in an antiferromagnet that is canted by an external magnetic field. In both cases,
the electron dynamics is obtained from nonequilibrium dynamical mean-field theory. We find that the modified $J_\mathrm{ex}$ emerges already within a few electron hopping times after the pulse, with a reduction that is comparable to the effect of chemical doping.
\end{abstract}



\maketitle
Magnetic long-range order and the dynamics of spins in magnetic materials are governed by the exchange interaction $J_\mathrm{ex}$, the strongest force of magnetism. Because $J_\mathrm{ex}$ emerges from the Pauli principle and the electrostatic Coulomb repulsion between electrons, it is sensitive to purely nonmagnetic perturbations. This fact implies intriguing and largely unexplored possibilities for the ultrafast control of magnetism by femtosecond laser pulses, which is currently a very active research area \cite{Kirilyuk10}. In principle, laser-excitation can effect $J_\mathrm{ex}$ by modulating the electronic structure (electron hopping, Coulomb repulsion) and by creating a nonequilibrium distribution of photoexcited carriers (photodoping). A modification of $J_\mathrm{ex}$ has been discussed within the context of experiments on manganites \cite{wall2009,Forst11,Li13}, magnetic semi-conductors \cite{Matsubara13}, and, using static field gradients, ultracold atoms in optical lattices \cite{duan2003,Trotzky2008}. While it might play a role as well in metallic ferromagnets \cite{ju2004,thiele2004,rhie2003,carley2012}, ultrafast demagnetization \cite{Beaurepaire96} and laser-induced magnetization reversal \cite{Stanciu07,Radu11,Ostler12} seem at least partly understood in terms of a given time-independent $J_\text{ex}$. Clearly, more theoretical work is needed to understand how {\em effective} a modification of $J_\mathrm{ex}$ under nonequilibrium conditions can be, and how {\em fast} $J_\mathrm{ex}$ can be modified. The latter touches the fundamental question for the time scale at which the description of spin dynamics in terms of a $J_\mathrm{ex}$ emerges from the full electronic dynamics, before which $J_\mathrm{ex}$ is not a valid concept at all. Although this question has not been directly addressed in the experiments mentioned above, an investigation of this ultimate limit of spin dynamics is in range using today's femtosecond laser technology.

In general, the exchange interaction arises from a low-energy description of the electronic states in terms of magnetic degrees of freedom. Recently, Secchi {\em et al}. defined the nonequilibrium exchange interaction via an effective action that governs the spin dynamics out of equilibrium, leading to an expression in terms of nonequilibrium electronic Green's functions \cite{Secchi13}. Here, we apply this framework to the paradigm single-band Mott-Hubbard insulator at half-filling, for which the concept of exchange interaction in equilibrium is very well understood. To directly assess the nonequilibrium electron dynamics and evaluate the nonequilibrium Green's functions, we employ nonequilibrium dynamical mean field theory (DMFT). Previous investigations of the antiferromagnetic phase in the Hubbard model have demonstrated ultrafast melting of long-range order after an interaction quench \cite{Werner2012afm,Tsuji2012}. Here, we will focus on the excitation with an electric field pulse and weaker excitation strength, to assess the control of $J_\mathrm{ex}$ within the ordered phase and to determine how fast a rigid spin dynamics emerges after the excitation.

{\em Model.---}
In this work we study the antiferromagnetic phase of the repulsive Hubbard model at
half-filling,
\begin{align}
\label{repulsive hubbard}
H
=&
-t_0\sum_{\langle ij \rangle \sigma}
c_{i\sigma}^\dagger
c_{j\sigma}
+
U
\sum_{j}
n_{j\uparrow}n_{j\downarrow}
+
B_x
\sum_j
S_{jx}.
\end{align}
Here $c_{i\sigma}^\dagger$ creates an electron at site $i$ with spin $\sigma=\su,\sd$ along a given spin quantization axis (the $z$ axis).  The first two terms describe nearest-neighbor hopping $t_0$ and repulsive on-site interaction $U$. The third term introduces coupling of the spin $S_{j\alpha}=\frac{1}{2} \sum_{\sigma\sigma'} c_{j\sigma}^\dagger (\hat \sigma_\alpha)_{\sigma\sigma'} c_{j\sigma'}$ to a homogeneous magnetic field $B_x$ along the $x$ axis ($\alpha=x,y,z$; $\hat \sigma_\alpha$ denote the Pauli matrices). The latter allows us to probe transverse dynamics of the antiferromagnetic order parameter in the $y$-$z$ plane; the $x$ component of the total spin $\langle S_x\rangle $ is conserved.

To solve the electron dynamics in the Hubbard model we use nonequilibrium DMFT \cite{Freericks2006,REVIEW}. Within DMFT \cite{Georges96}, which becomes exact in the limit of infinite dimensions \cite{Metzner1989}, local correlation functions are obtained from an effective impurity model in which one site of the lattice is coupled to a noninteracting bath. In the presence of a transverse magnetic field $B_x$ one must include spin-flip terms in the effective impurity action, which thus takes the form $\mathcal{S}=\mathcal{S}_\text{loc}-\text{i}\!\int\!dt\!\int\!dt' \sum_{\sigma\sigma'} c_{\sigma}(t)^\dagger \Delta_{\sigma\sigma'}(t,t') c_{\sigma'}(t')$. Here, $\Delta_{\sigma\sigma'}(t,t')$ is the hybridization function of the bath that is determined self-consistently. The impurity model is solved within the perturbative hybridization expansion (noncrossing approximation, NCA). The incorporation of spin-flip terms $\Delta_{\uparrow\downarrow}$ is a straightforward extension to the nonequilibrium DMFT implementation and the NCA, which have been explained in Refs.~\cite{REVIEW} and \cite{Eckstein2010nca}. For completeness, we summarize explicit equations in the Supplementary Material. In general, the DMFT approximation is expected to be appropriate when local correlations dominate, such as is the case in the Mott-insulating phase for the short-time dynamics (up to $\sim$100 fs), when the much slower ($\sim$ps and beyond) inhomogeneous dynamics (spin waves, domain growth) is not yet developed. The reliability of the NCA impurity solver has been tested in equilibrium and for short-time dynamics by comparison with higher-order hybridization expansions as well as with the numerically exact quantum Monte Carlo impurity solver. Good agreement was found at large $U$ in the paramagnetic phase \cite{Eckstein2010nca,Eckstein2012c} and for the antiferromagnetic phase boundary \cite{Werner2012afm}.

{\em Nonequilibrium exchange interactions.---}
For general nonequilibrium situations, the exchange interaction is defined in terms of
an effective spin action that reproduces the spin dynamics of the full electronic model. A formal derivation of the spin interaction in such a model has been given by Secchi and co-workers \cite{Secchi13}. The essential idea is to define the effective spin action in terms of time-dependent rotations of the spin quantization axes $\mathbf{e}_i(t)$, as described by Holstein-Primakov bosons $\xi_i(t)$. Starting from the electronic partition function as a path integral over fermionic fields $\phi$, one introduces rotated fermion fields $\psi$ and then expands the action to second order in $\xi$. 
The rotated fermionic fields are integrated out, which leads to spin action with an interaction term of the form $\mathcal{S}_\text{spin}[\xi^*\!,\xi]=\sum_{ij}\int\!dt\!\int\!d\tp\,\xi_{i}^*(t)A_{ij}(t,t')\xi_{j}(\tp)$. The coupling $A_{ij}(t,\tp)$ between spin rotations at different times 
and different sites $i\neq j$ is expressed in terms of the 
spin-dependent single-particle Green's functions 
$G^\sigma_{ij}(t,\tp)$ and the self-energies $\mathit{\Sigma}^\sigma_{ij}(t,\tp)$,
\begin{eqnarray}
A_{ij}(t,\tp) &=& R^\sd_{ij}(t,\tp)R^\su_{ji}(\tp,t) + S_{ij}^\sd(t,\tp)S_{ji}^\su(\tp,t) \nonumber \\
&& - T^\sd_{ij}(t,\tp)G^{\su}_{ji}(\tp,t) - G^{\sd}_{ij}(\tp,t) T^\su_{ji}(\tp,t),
\label{jex:A}
\end{eqnarray}
where $T^\sigma_{ij}(t,\tp)=\mathit{\Sigma}^\sigma_{ij}(t,\tp)+\left[ \mathit{\Sigma} \cdot G \cdot \mathit{\Sigma} \right]^\sigma_{ij}(t,\tp)$, $R^\sigma_{ij}(t,\tp)=\left[ G \cdot \mathit{\Sigma} \right]^\sigma_{ij}(t,\tp)$, and $S^\sigma_{ij}(t,\tp)=\left[ \mathit{\Sigma} \cdot G \right]^\sigma_{ij}(t,\tp)$. These formulas are a direct generalization of the equilibrium formalism \cite{Katsnelson00,Katsnelson02}, 
which is based on variations of the total (free) energy $\delta E=J_\text{ex}\theta^2$
for static spin rotations by a small angle $\theta$. We emphasize that  Eq.~\eqref{jex:A} is valid for arbitrary fast and strong fields, apart from neglecting of vertex corrections \cite{Secchi13}. In addition, the expressions assume rotations from a collinear state. Reduction of the action with a retarded (two-time) exchange coupling to a spin {\em Hamiltonian} with an instantaneous (possibly time-dependent) interaction is possible when the rotations of the quantization axes are much slower than the electron dynamics, and, in particular, slower than time-dependent fluctuations of the local magnetic moments themselves. Then, we can average over the fast electron dynamics, 
\begin{equation}
\label{jex:integral}
J_{ij}(t)=\mathrm{Im} \int_0^\infty \!ds A^\mathrm{ret}_{ij}(t,t-s).
\end{equation}
Still, $J_{ij}(t)$ contains not only the exchange interactions, but also the time-averaged 
reduction of the local spin by fluctuations. The "bare" exchange interactions between 
spin vectors $\langle\mathbf{S}_i\rangle$ are finally given by
\begin{equation}\label{e:jexbare}
J^0_{ij}(t)=\frac{1}{4}\frac{J_{ij}(t)}{\langle{S}_{iz}\rangle\langle{S}_{jz}\rangle}.
\end{equation}
In the regime where a rigid macrospin model is valid, $J_{ij}^0$ should determine the spin dynamics by a Landau-Lifshitz equation. For a canted antiferromagnet on a bipartite lattice in a transverse magnetic field $B_x$, we can write $\langle \dot{\mathbf{S}}_1\rangle = -\langle \mathbf{S}_1\rangle\times\mathbf{B}_\text{eff}$, where $\mathbf{B}_\text{eff} = 2J_\text{ex}\langle\mathbf{S}_2\rangle+B_x\mathbf{e}_x$. Here $\langle \mathbf{S}_{1,2}\rangle$ correspond to the spin on the two sublattices, and the effective exchange interaction is $J_\text{ex}=\sum_{j} J_{ij}^0$.
Using N\'eel symmetry $\langle S_{1y,z}\rangle=-\langle S_{2y,z}\rangle$, 
$\langle S_{1x}\rangle=+\langle S_{2x}\rangle$ we can infer the exchange interaction in the canted geometry from the spin dynamics,
\begin{equation}\label{e:jexc}
J^c_\text{ex}=-\frac{B_x}{4\langle S_{1x}\rangle} - \frac{1}{4\langle S_{1x}\rangle}\frac{\langle \dot{S}_{1y}\rangle}{\langle S_{1z}\rangle }.
\end{equation}
The validity of the instantaneous approximation is a fundamental question that is not resolved in general, and which will be partially addressed below by comparison of the two Eqs. \eqref{e:jexbare} and \eqref{e:jexc}.

{\em Results.---}
We first solve the DMFT equations on the Bethe lattice with a semielliptic density of states $D(\epsilon)=\sqrt{4-\epsilon^2}/2\pi$. This setup implies a closed-form self-consistency condition and allows us to compute the electronic dynamics to long times, as needed for an accurate evaluation of the integral in Eq.~\eqref{jex:integral} (see the Supplementary Material).

\begin{figure}[tbp]
\includegraphics[width=\columnwidth]{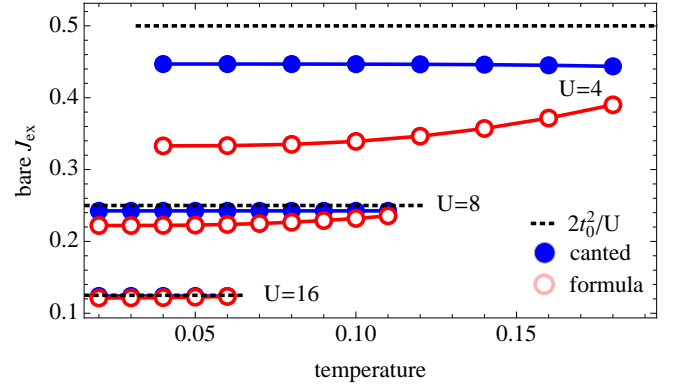}
\caption{Bare exchange interaction as function of temperature for different values $U$, computed from the formula Eq.~(\ref{e:jexbare}) (red open circles) and from the canted geometry Eq.~(\ref{e:jexc}) (blue solid discs). For large $U$ the calculations show excellent agreement with the analytical result $|J^a_\mathrm{ex}|=2t_0^2/U$ indicated with dashed lines.
\label{fig1}}
\end{figure}

Before exploring nonequilibrium, it is illustrative to evaluate the exchange interaction \eqref{e:jexbare} in the familiar equilibrium case. For the Mott insulator at half-filling, the static exchange interaction at zero temperature can be obtained from a perturbation expansion in the hopping, which yields $|J_\mathrm{ex}^a|=2t_0^2/U$. In Fig.~\ref{fig1}, we compare the  analytical value $|J_\mathrm{ex}^a|$ (dashed lines) and the bare exchange interaction $|J_\mathrm{ex}^0|=|J_\mathrm{12}^0|$ computed from the collinear DMFT solution using Eq.~\eqref{e:jexbare} (red circles) as function of temperature for three values of $U$. In addition, we solve the DMFT equations for the antiferromagnetic Mott insulator in a weak transverse field of strength $B_x$ and obtain an estimate $|J_\mathrm{ex}^c|=|B_x/4\langle S_x\rangle|$ by comparing the canting $\langle S_x\rangle$ of spins to the prediction from a rigid macrospin model Eq.~\eqref{e:jexc} in the static limit (blue solid disks). We choose $B_x=0.64t_0^2/U$, such that the canting angle at low temperature is about 10 degrees for all $U$. At large $U$, we find excellent agreement between $J_\mathrm{ex}^a$, $J_\mathrm{ex}^{0}$, and $J_\mathrm{ex}^c$, where deviations between $J_\mathrm{ex}^c$ and $J_\mathrm{ex}^a$ are on the order of $(t_0/U)^2$, which also confirms the validity of the DMFT approximation for studying exchange interactions. For smaller $U$, the deviation of $J_\mathrm{ex}^{0}$ from $J_\mathrm{ex}^c$ becomes more pronounced, up to 25\% at $U=4$. The differences between the two Eqs.~\eqref{e:jexbare} and \eqref{e:jexc} may have several possible origins: (i) At small values of $U$, the rigid macrospin model is no longer valid, because retardation effects in $A(t,t')$ become relevant, (ii) vertex corrections to Eq.~\eqref{jex:A} become important, or (iii) $J_\mathrm{ex}^{0}$ is a nearest-neighbor interaction while Eq.~\eqref{e:jexc} also takes into account next-nearest-neighbor terms. Below we will study nonequilibrium exchange at large values of $U$. Nevertheless, for moderate $U$, where retardation effects to the exchange become important, we can still use Eq.~\eqref{e:jexc} as a heuristic measure for $J_\text{ex}$, in the sense that it is the best estimate of an instantaneous exchange interaction which is in accordance with an observed spin dynamics.

\begin{figure}[t]
\includegraphics[width=\columnwidth]{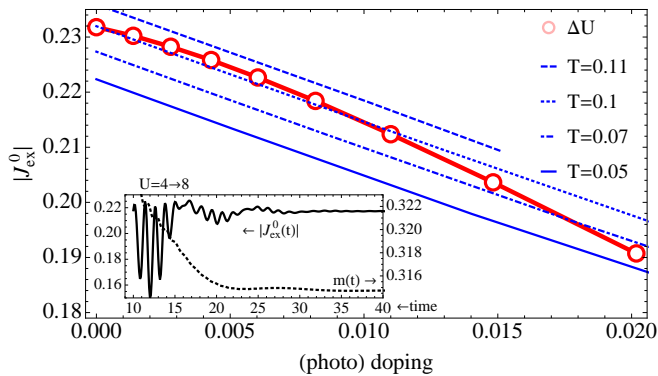}%
\caption{
Comparison of the nonequilibrium exchange interaction (open circles) in the quasistationary state after an interaction quench $\Delta U$ in the Bethe lattice with the equilibrium exchange interaction of the chemically doped model (solid symbols) for $U=8$ and different temperatures. The inset shows the bare time-dependent exchange interaction (solid line) and staggered magnetization (dashed line) caused by the quench $U=4\rightarrow 8$. \label{fig2}}
\end{figure}

Next, we investigate how \textit{fast} $J_\text{ex}$ can be modified under electronic nonequilibrium conditions, which we generate by suddenly changing $U$. It was recently demonstrated that after such an interaction quench the order parameter $m$ quickly relaxes to a quasistationary but nonthermal value \cite{Werner2012afm} that is protected from further decay by the slow recombination rate of doublons and holes \cite{Eckstein11,Sensarma2010a,Lenarcic2012preprint,Moritz2012}. This transient
state resembles properties of a photodoped system in which charge carriers are created by a short laser pulse. We will refer to the induced change of the doublon and hole densities $d$ and $h$ with respect to their equilibrium values $d_0$ and $h_0$ as photodoping $\Delta n=d+h-d_0-h_0=2(d-d_0)$. The inset of Fig.~\ref{fig2} shows the evolution of the time-dependent nonequilibrium exchange interaction (solid line) and order parameter (dashed line) for a quench $U=4\rightarrow8$. [A Gaussian window $\exp(-s^2/w^2)$ of length $w=10t_0/\pi$ was used in Eq.~\eqref{jex:integral} to ensure a smooth cut off of the upper integration limit.] We find that $|J_\mathrm{ex}^0|$, like $m$, becomes stationary already on an electronic time scale, which shows the emergence of a spin Hamiltonian on the timescale of a few tens of inverse hoppings.

To study how \textit{effective} $J_\mathrm{ex}$ is modified, we evaluate it in the quasistationary state after different excitation strengths $\Delta U = U_\text{f} - U_\text{i}=0,\ldots,4$, with final $U_\text{f}=8$. The result is shown by red open circles in Fig.~\ref{fig2} as a function of "photo-doping" $\Delta n$, demonstrating a reduction of $J_\mathrm{ex}$ to a value significantly below the equilibrium difference $J_\text{ex}^0(U_\text{i})-J_\text{ex}^0(U_\text{f})$. The results are independent of a Gaussian cutoff in Eq.~\eqref{jex:integral} for $w=60t_0/\pi$. Only for the largest $\Delta U$ do we find a slight dependence on $w$ that indicates that $J_\mathrm{ex}^0$ is not yet fully stationary. Furthermore, the blue lines in Fig.~\ref{fig2} show the equilibrium exchange interaction at chemical doping for different temperatures. These results confirm the conclusions obtained from analyzing the electronic spectrum \cite{Werner2012afm}, that properties of the photodoped state with added doublons and holes resemble those of the chemically doped state with the same total number of carriers: Adding doublons and holes causes an ultrafast weakening, or ``quenching'' of  the exchange interaction by an amount comparable to that of chemical doping. Qualitatively, the weakening of the antiferromagnetic exchange can result from a lowering of the kinetic energy of mobile carriers in a parallel spin alignment (for $U=\infty$ and small doping ferromagnetism is favored \cite{Nagaoka66}). 

\begin{figure}[ht]
\includegraphics[width=\columnwidth]{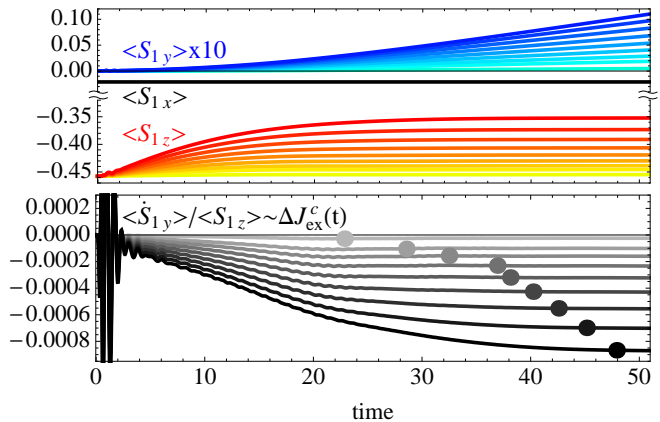}
\caption{Induced spin dynamics (top) and modification of the exchange interaction (bottom) caused by excitation 
with an electric field (hypercubic lattice, $U=8$).
\label{fig3}}
\end{figure}

\begin{figure}[ht]
\includegraphics[width=\columnwidth]{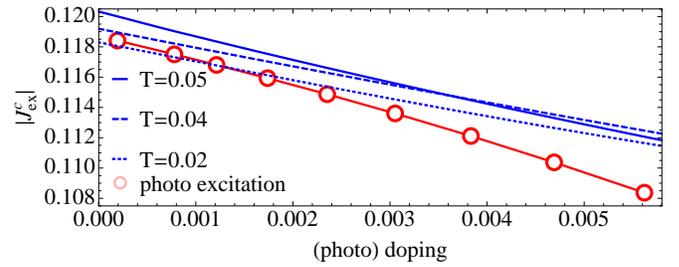}
\caption{Comparison of the nonequilibrium exchange interaction (red open circles) computed from the induced precession (Fig.~\ref{fig3}), with the equilibrium exchange interaction in the chemically doped system (blue lines).
\label{fig4}}
\end{figure}

{\em Photoexcitation.---}
To further demonstrate the possibility of changing $J_\mathrm{ex}$ in a setup that is closer to the laser excitation of condensed-matter systems, we study the Hubbard model driven by an external electric field. This is implemented for the infinite-dimensional hypercubic lattice with density of states $D(\epsilon)=\exp(-\epsilon^2)/\sqrt{\pi}$, with the electric field pointing along the body diagonal \cite{REVIEW,Turkowski2005a}. Photoexcited carriers are created by a single-cycle pulse $E(t) = E_0 \sin(\omega t ) \exp[-\alpha(t-t_c)^2]$, $t_c=\pi/\omega$, $\alpha=4.6/t_c^2$ with a Gaussian envelope and a center frequency of $\omega=U$. To directly measure the transverse spin dynamics associated with $J_\text{ex}$, we study the system in a canted geometry induced by a homogeneous magnetic field $B_x$. Before laser excitation, the system is prepared in equilibrium with a canting angle determined by the balance of $B_x$ and $J_\text{ex}$. When $J_\text{ex}$ is changed, this balance will be broken and a spin resonance will be excited. Such spin resonances can, in principle, be detected experimentally using magneto-optical techniques \cite{Kirilyuk10} and THz spectroscopy \cite{Nishitani2011}. In our simulations, we extract the nonequilibrium exchange interaction by comparing the spin dynamics obtained within DMFT to the rigid macrospin model, cf. Eq.~\eqref{e:jexc}. The results of this approach are shown in Fig.~\ref{fig3}, computed at $U=8$, $B_x=0.01$, and initial temperature $T=0.03$. The top panel shows that the sublattice magnetization is initially in the $x$-$z$ plane. Light to dark colors indicate excitation strengths ranging from $|E_0|/t_0=1$ to $5.5$. The bottom panel shows $\Delta J_\mathrm{ex}^0\sim \langle \dot{S}_{1y}\rangle/\langle S_{1z}\rangle$, cf. Eq.~\eqref{e:jexc}, where $\langle \dot{S}_{1y}\rangle$ is computed from the time trace of $\langle S_{1y}(t)\rangle$. We observe three different time scales in our simulations: (i) Fast $1/U$ oscillations on the timescale of the laser excitation, as most clearly seen in the bottom panel. This characterizes the stabilization of the local magnetic moments. (ii) Relaxation of the order parameter and the exchange interaction. (iii) The onset of rigid rotation of the spin sublattices at quasistationary values $|\langle\mathbf{S}_1\rangle|$ and $J^c_\text{ex}$. We estimate the time $t_*$ that it takes for $J^c_\text{ex}$ to become stationary from $J^c_\text{ex}(t_*)-J^c_\text{ex}(t_\text{max})<\varepsilon$, where $\varepsilon$ is the numerical accuracy. The values $t_*$,  which are indicated as dots in the bottom panel of Fig.~\ref{fig3}, show that a quasistationary state and rigid spin dynamics emerge after a few tens of inverse hoppings, similar as for the sudden change of $U$. This relaxation time increases with the excitation density, as the critical excitation for melting the antiferromagnetic order is approached, but is much shorter than the period of spin precession in the field, which supports the interpretation that photoexcitation causes an ultrafast quenching of $J_\mathrm{ex}$. Furthermore, we find that direct photoexcitation has a similar effect as the interaction quench; \textit{i.e.}, the efficiency of the modification of $J_\text{ex}$ is determined by the number of photoexcited carriers. This is demonstrated in Fig.~\ref{fig4} by plotting the extracted exchange interaction in the quasistationary state as a function of the photodoping, together with equilibrium calculations in the canted geometry with chemical doping. In the hypercubic lattice, we observe that photoexcitation modifies $J_\textrm{ex}^c$ slightly stronger than chemical doping. In addition, there is a more pronounced temperature dependence of $J_\textrm{ex}^c$ in equilibrium. Both effects might be related to a slightly different dynamics of low-energy (photo-) doped carriers in the Bethe lattice and the hypercubic lattice, where the latter does not have a sharp band edge in the density of states.

In summary, we report that photoexcitation causes an ultrafast quenching of the exchange interaction in a Mott insulator. An effectively static $J_\mathrm{ex}$ can be defined already on the ultrafast time scale on the order of a few tens of inverse hopping times, which is similar to the relaxation time of the order parameter. The reduction of $J_\mathrm{ex}$ is comparable to that of a chemically doped state when measured in terms of the total number of excited carriers. These results demonstrate intriguing possibilities to control magnetic order without magnetic fields. Similar, or even more efficient ways to control $J_\text{ex}$ under nonequilibrium conditions might be found by extending our work to more complex multi-band systems such as the prototype Mott-insulator V$_2$O$_3$ \cite{Liu2011} and to materials with different exchange mechanisms.

\begin{acknowledgments}
We thank K. Balzer, S. Brener, A. Secchi, M.I. Katsnelson, A.V. Kimel, J. Kroha, A. Lichtenstein and Ph. Werner for fruitful discussions. The calculations were run on the supercomputer HLRN-II of the North-German Supercomputing Alliance. J.H.M. acknowledges funding from the Nederlandse Organisatie voor Wetenschappelijk onderzoek (NWO Rubicon-grant).
\end{acknowledgments}

\appendix

\begin{center}
{\large{\bf Supplementary material}}
\end{center}
{\center}

\section{Implementation of nonequilibrium DMFT with a transverse magnetic field}

In the supplementary material we describe in detail how we implement the nonequilibrium 
DMFT for the antiferromagnetic phase of the Hubbard model in a transverse magnetic field,
\begin{align}
\label{repulsive hubbard}
H
=&
-\sum_{\langle ij \rangle \sigma}
t_{ij}\,
c_{i\sigma}^\dagger
c_{j\sigma}
+
U
\sum_{j}
n_{j\uparrow}n_{j\downarrow}
+
B_x
\sum_j
S_{jx}
-\mu \sum_{j\sigma} n_{j\sigma}.
\end{align}
Apart from the incorporation of spin-flip terms in the DMFT impurity action and 
the self-consistency relations, the resulting equations are a straightforward 
extension of the equations for the paramagnetic phase and the collinear 
antiferromagnet, which have been described previously \cite{REVIEW,Eckstein2010nca}. 

\subsection{ The impurity model}

To describe a magnetically ordered system, we introduce  Keldysh Green's functions 
which are $2\times2$ matrices in spin-space,
\begin{align}
\label{gloc}
&
\hat G_{jj'}(t,t')
=
-i
\langle \,
\Tc
\hat \spinor_j(t) \hat \spinor_{j'}^\dagger(t)
\, 
\rangle 
_{\mathcal{S}} 
\\
&=
\begin{pmatrix}
-i\langle 
\Tc
\,c_{j\uparrow}(t)\,c_{j'\uparrow}^\dagger(t)\, \rangle 
_{\mathcal{S}} &
-i\langle\Tc
 \,c_{j\uparrow}(t)\,c_{j'\downarrow}^\dagger(t)\,\rangle 
 _{\mathcal{S}} 
\\
-i\langle\Tc
 \,c_{j\downarrow}(t)\,c_{j'\uparrow}^\dagger(t)\,\rangle 
 _{\mathcal{S}} 
&
-i\langle\Tc
 \,c_{j\downarrow}(t)\,c_{j'\downarrow}^\dagger(t)\,\rangle 
 _{\mathcal{S}} 
\end{pmatrix}.
\end{align}
Here $\hat \spinor_j$ is the spinor 
\begin{equation}
\hat \spinor_j
= 
\begin{pmatrix}
c_{j\uparrow}
\\
c_{j\downarrow}
\end{pmatrix},
\end{equation}
$\CC$ is the $L$-shaped Keldysh 
contour that extends from $0$ to some maximum time $t_\text{max}$ along the real 
axis, back to $0$, and to $-i\beta$ along the imaginary time axis, and 
\begin{equation}
\langle 
\Tc
\cdots
\rangle
_{\mathcal{S}}
\equiv
\Tr \big[ \Tc e^{\mathcal{S}}\,  \cdots  \big] \,/\,\Tr \big[\Tc e^{\mathcal{S}}\,\big]
\end{equation}
denotes the contour-ordered expectation value for an action $\mathcal{S}$; the 
action for the lattice model \eqref{repulsive hubbard} is given by 
$\hat{\mathcal{S}}=-i \int_\CC d\bar t \,\hat H(\bar t) $. We follow Ref.~\cite{REVIEW} 
for the notation for Keldysh Green's and their convolution and time-derivatives along 
$\CC$. 

The antiferromagnetic DMFT solution is obtained on a bipartite lattice at and close 
to half-filling.  The local Green's function $\hat G_\alpha$ for a site on sub-lattice 
$\alpha=A,B$ of the bipartite lattice is obtained from an impurity model with action 
\begin{align}
\label{dmft action}
\hat{\mathcal{S}}_\alpha
=
-i\int_\CC
dt
\,H_{\text{loc},\alpha}(t)
-i
\int_\CC
\!dt dt'\,
\hat \spinor ^\dagger(t)
\hat \Delta_\alpha(t,t')
\hat \spinor(t'),
\end{align}
where $H_{\text{loc},\alpha}(t)$ is the local part of Hamiltonian \eqref{repulsive hubbard},
and $\hat \Delta_\alpha(t,t')$ is the hybridization matrix, which is later determined self-consistently.

In order to compute compute the impurity Green's function
\begin{align}
\label{gloc dmft}
\hat G_{\alpha}(t,t')
=
-i
\langle \,
\Tc
\hat \spinor(t) \hat \spinor ^\dagger(t)
\, 
\rangle 
_{\hat{\mathcal{S}}_\alpha},
\end{align}
we use the lowest strong-coupling impurity solver \cite{Eckstein2010nca} 
(non-crossing approximation, NCA), which is a self-consistent expansion in the 
hybridization function $\hat \Delta_\alpha(t,t')$. The hybridization expansion can be  
formulated in terms of pseudo-particle propagators $\pseudoG_{nm}$,  whose flavor 
indices $n$, $m$ correspond to the many-body states of the impurity. These propagators 
satisfy a Dyson equation, with a self-energy $\pseudoS$ that is given by a diagrammatic 
expansion in the hybridization function. In the present case, a basis of the local Hilbert 
space at the impurity model is given by the four states $|0\rangle$,  $|\sigma\rangle \equiv 
c_\sigma^\dagger |0\rangle$ (for $\sigma=\pm$), and $|2\rangle\equiv c_\uparrow^\dagger 
c_\downarrow^\dagger |0\rangle$. Because the $H_\text{loc}$ and $\hat \Delta_\alpha(t,t')$ 
allow spin-flip terms, pseudo-particle propagators $\pseudoG_{nm}$ are only diagonal in particle number of $|n\rangle$ and 
$|m\rangle$, but not in spin. Hence, we introduce three propagators,  $\pseudoG^{(0)} \equiv \pseudoG_{|0\rangle,|0\rangle}$, 
$\pseudoG^{(2)} \equiv \pseudoG_{|2\rangle,|2\rangle}$, and  $\pseudoG^{(1)}_{\sigma\sigma'} 
\equiv \pseudoG_{|\sigma\rangle,|\sigma'\rangle}$, and corresponding self-energies $\pseudoS^{(n)}(t,t')$, 
$n=0,1,2$. From the the diagrammatic rules for a general multi-orbital case as stated in 
Ref.~\cite{Eckstein2010nca}, we obtain
\begin{align}
\pseudoS^{(0)}(t,t')
&=
-i\sum_{\sigma,\sigma'}
\pseudoG ^{(1)}_{\sigma\sigma'}(t,t')
\Delta_{\sigma'\sigma}(t',t)
\\
\pseudoS^{(1)}_{\sigma\sigma'}(t,t')
&=
i
\pseudoG ^{(0)}(t,t')
\Delta_{\sigma\sigma'}(t,t')
-i \bar\sigma \bar\sigma'
\pseudoG ^{(2)}(t,t')
\Delta_{\bar\sigma' \bar \sigma}(t',t)
\\
\pseudoS^{(2)}(t,t')
&=
i
\sum_{\sigma\sigma'}
\bar \sigma \bar \sigma'
\pseudoG ^{(1)}_{\sigma\sigma'}(t,t')
\Delta_{\bar \sigma \bar \sigma'}(t,t').
\end{align}
Finally, the local Green's function is given by evaluation of the ``bubble diagram'' \cite{Eckstein2010nca}
\begin{align}
G_{\sigma\sigma'}(t,t')
&=
i
\pseudoG ^{(0)}(t',t)
\pseudoG ^{(1)}_{\sigma\sigma'}(t,t')
-i
 \sigma \sigma'
\pseudoG ^{(1)}_{\bar\sigma'\bar\sigma}(t',t)
\pseudoG ^{(2)}(t,t').
\end{align}
 

\subsection{DMFT self-consistency for the Bethe lattice}

For a Bethe lattice with nearest neighbor hopping  $t_{ij} \equiv t_0/\sqrt{Z}$ in the limit $Z\to\infty$,
which has a semi-elliptic density of states $D(\epsilon)=\sqrt{4-\epsilon^2}/2\pi$ for $t_0=1$, the 
hybridization function is  determined by the closed form self-consistency equation \cite{Georges96}
\begin{equation}
\label{bethe}
\hat \Delta_\alpha(t,t')
=
t_0^2 \hat G_{\bar \alpha}(t,t'),
\end{equation}
where $\bar \alpha=B$($A$) for $\alpha=A$($B$).
The NCA equations together with Eq.~\eqref{bethe} provide a closed set of equations 
that is numerically propagated in time as described in Ref.~\cite{REVIEW}. 


\subsection{DMFT self-consistency for the hypercubic lattice}

To solve the DMFT equation on a cubic lattice with $A/B$ sub-lattice symmetry breaking, we let  $\lattC$ 
denote the magnetic superlattice of points $\bm R_j$, and $\latt$ is the full lattice with atoms at coordinates 
$\bm r_{j\alpha} = \bm R_j+\bm \delta_\alpha$, $\alpha=A,B$. For example, we can choose $\lattC$ as the 
$A$-sublattice of the bipartite cubic lattice, such that $\bm \delta_A=0$, $\bm \delta_B=(1,0,...)$. 
We then introduce the Fourier transform with respect to the coordinate $\bm R_j$,
\begin{align}
\label{ck cluster}
c_{\KK\alpha\sigma}
&=
\frac{1}{\sqrt{L'}}
\sum_{j\in \lattC} 
e^{-i\KK ({\bm R}_j+\bm \delta_\alpha)}
c_{j\alpha\sigma},
\\
c_{j\alpha\sigma}
&=
\frac{1}{\sqrt{L'}}
\sum_{\KK\in \bzC} 
e^{i\KK ({\bm R}_j+\bm \delta_\alpha)}
c_{\KK\alpha\sigma},
\end{align}
where $L'$ is the number points in $\lattC$,
and $\bzC$ is the first Brillouin zone of the magnetic superlattice. To describe 
the broken symmetry phase, we introduce super-spinors
\begin{align}
\hat \sspinor_{\KK}=
\begin{pmatrix}
c_{\KK,A,\uparrow}
\\
c_{\KK,A,\downarrow}
\\
c_{\KK,B,\uparrow}
\\
c_{\KK,B,\downarrow}
\end{pmatrix}
\equiv
\begin{pmatrix}
\hat \spinor_{\KK,A}
\\
\hat \spinor_{\KK,B}
\end{pmatrix},
\end{align}
and corresponding Green's functions
\begin{align}
\hat G_\KK (t)
&=
-i \langle 
\mathcal{T}_\CC
\hat \sspinor_\KK(t)
\hat \sspinor_\KK^\dagger(t')
\rangle
\\
&\equiv
\begin{pmatrix}
\hat G_{\KK,AA}(t,t') 
&
\hat G_{\KK,AB}(t,t') 
\\
\hat G_{\KK,BA}(t,t') 
&
\hat G_{\KK,BB}(t,t') 
\end{pmatrix},
\end{align}
where the second expression is a block-matrix with entries
$\hat G_{\KK,\alpha\alpha'}(t,t') = -i \langle \mathcal{T}_\CC
\hat \spinor_{\KK,\alpha}(t)
\hat \spinor_{\KK,\alpha'}^\dagger(t')
\rangle$.
With this, the quadratic part of the Hamiltonian \eqref{repulsive hubbard} can be rewritten as
$\sum_{\KK\in \bzC} \hat \sspinor_{\KK}^\dagger \hat H_\KK \hat \sspinor_{\KK}$, with
\begin{align}
\hat H_{\KK}
=
\begin{pmatrix}
\hat H_{loc,A} 
& 
\hat \epsilon_\KK
\\
\hat \epsilon_\KK
& 
\hat H_{loc,B} 
\end{pmatrix}
,
\end{align}
where $\hat H_{loc,A}=\hat H_{loc,B} =\hat \sigma_x B_x$ and $\hat \epsilon_\KK = \epsilon_\KK\hat 1$,
with the $2\times 2$ unit matrix $\hat 1$. The electronic dispersion $\epsilon_\KK$ 
may be time-dependent due to inclusion of a external electric field via the Peierls
substitution (see below). The Dyson equation has a $4\!\times\!4$-structure,
\begin{align}
\label{latt inv 4}
\hat G_\KK^{-1}(t,t') = 
\delta_\CC(t,t') [(i\partial_t + \mu) \hat 1- \hat H_\KK ]
- \hat \Sigma(t,t'),
\end{align}
with the spatially local self energy 
\begin{align}
\hat \Sigma(t,t')
&=
\begin{pmatrix}
\hat \Sigma_{A}(t,t')
&
0
\\
0&\hat \Sigma_{B}(t,t')
\end{pmatrix},
\\
\hat \Sigma_\alpha(t,t')
&=
\begin{pmatrix}
\Sigma_{\alpha,\uparrow\uparrow}(t,t')
&
\Sigma_{\alpha,\uparrow\downarrow}(t,t')
\\
\Sigma_{\alpha,\downarrow\uparrow}(t,t')
&
\Sigma_{\alpha,\downarrow\downarrow}(t,t')
\end{pmatrix}.
\end{align}

Numerically, it is convenient to solve the DMFT equations without explicitly solving for the 
self energy. By introducing  $\hat Z_\alpha =  [i\partial_t + \mu - \hat H_{loc,\alpha}-\hat \Sigma_\alpha]^{-1}$, 
the impurity Dyson equation reads
\begin{align}
\label{impdyson}
\hat G_\alpha = \hat Z_\alpha + \hat Z_\alpha \conv \hat \Delta_\alpha \conv \hat G_\alpha,
\end{align}
and the lattice Dyson equation is given by
\begin{align}
\label{latt dyson z}
\hat G_\KK = 
\begin{pmatrix}
\hat Z_A & 0 
\\
0 & \hat Z_B
\end{pmatrix}
+
\begin{pmatrix}
\hat Z_A & 0 
\\
0 & \hat Z_B
\end{pmatrix}
\conv
\begin{pmatrix}
0 & \hat \epsilon_\KK 
\\
\hat \epsilon_\KK  & 0
\end{pmatrix}
\conv
\hat G_\KK.
\end{align}
The lattice Dyson equation can be written explicitly for its four $2\times2$ components,
\begin{align}
\hat G_{\KK,AA} 
&= 
\hat Z_A 
+
\hat Z_A 
\conv 
\hat \epsilon_\KK
\conv 
\hat Z_B
\conv \hat \epsilon_\KK
\conv 
\hat G_{\KK,AA},
\\
\hat G_{\KK,BB} 
&= 
\hat Z_B 
+
\hat Z_B 
\conv 
\hat \epsilon_\KK
\conv 
\hat Z_A
\conv 
\hat \epsilon_\KK
\conv 
\hat G_{\KK,BB},
\\
\label{gab}
\hat G_{\KK,AB} 
&= 
\hat Z_A 
\conv \hat \epsilon_\KK
\conv 
\hat G_{\KK,BB},
\\
\label{gba}
\hat G_{\KK,BA} 
&= 
\hat Z_B 
\conv \hat \epsilon_\KK
\conv 
\hat G_{\KK,AA},
\end{align}
(where we have reinserted the expressions for $\hat G_{\KK,AB} $ and $\hat G_{\KK,BA} $ into the 
equations for $\hat G_{\KK,AA}$ and $\hat G_{\KK,BB} $.) By summing these equations over $\KK$ 
and comparing with the impurity Dyson equation, we then obtain an explicit equation for the hybridization 
function (for $\alpha=A,B$). For this it is convenient to introduce the moments
\begin{align}
\hat G_\alpha 
&= \frac{1}{L'}\sum_{\KK\in \bzC} \hat G_{\KK,\alpha\alpha}
\label{g-1}
\\
\hat G_\alpha^{(1)} 
&\equiv 
\frac{1}{L'}\sum_{\KK\in \bzC} 
\hat \epsilon_\KK
\conv 
\hat Z_{\bar \alpha}
\conv 
\hat \epsilon_\KK
\conv 
\hat G_{\KK,\alpha\alpha}
\label{g1-1}
\\
&=
\hat \Delta_\alpha \conv G_\alpha,
\label{g1-2}
\\
\hat G_\alpha^{(2)} 
&\equiv 
\frac{1}{L'}\sum_{\KK\in \bzC} 
\hat \epsilon_\KK
\conv 
\hat Z_{\bar \alpha}
\conv 
\hat \epsilon_\KK
\conv 
\hat G_{\KK,\alpha\alpha}
\conv 
\hat \epsilon_\KK
\conv 
\hat Z_{\bar \alpha}
\conv 
\hat \epsilon_\KK
\nonumber\\
&+
\frac{1}{L'}\sum_{\KK\in \bzC} 
\hat \epsilon_\KK
\conv 
\hat Z_{\bar \alpha}
\conv 
\hat \epsilon_\KK
\label{g2-1}
\\
&=
\hat \Delta_\alpha+\hat \Delta_\alpha \conv G_\alpha\conv \hat \Delta_\alpha.
\label{g2-2}
\end{align}
Here Eqs.~\eqref{g1-2} and \eqref{g2-2} follow from comparison with the impurity Dyson equation 
\eqref{impdyson}. Combining the two equations, we obtain
\begin{align}
\label{get delta}
(\hat 1 + \hat G_\alpha^{(1)}) \conv \hat \Delta_\alpha
=
\hat G^{(2)}_\alpha,
\end{align}
from which the hybridization can be determined, thus closing the DMFT self-consistency.

Throughout this work we consider magnetic fields along $x$, perpendicular to the antiferromagnetic 
order parameter. In this case, the system is invariant under a translation by one lattice constant and 
spin rotation by $\pi$ around the axis of the $B$-field. This symmetry can be used to relate local 
quantities at the $A$ and $B$ sites, i.e.,
\begin{align}
\label{symmetry}
\hat \Sigma_B
&=
\hat \sigma_x
\hat \Sigma_A
\hat \sigma_x,
\end{align}
and analogous for the functions $\hat Z_\alpha$, $\hat G_\alpha$, and $\hat \Delta_\alpha$.
Explicitly,
\begin{align}
\begin{pmatrix}
\Sigma_{B,\uparrow\uparrow}(t,t')
&
\Sigma_{B,\uparrow\downarrow}(t,t')
\\
\Sigma_{B,\downarrow\uparrow}(t,t')
&
\Sigma_{B,\downarrow\downarrow}(t,t')
\end{pmatrix}
&=
\begin{pmatrix}
\Sigma_{A,\downarrow\downarrow}(t,t')
&
\Sigma_{A,\downarrow\uparrow}(t,t')
\\
\Sigma_{A,\uparrow\downarrow}(t,t')
&
\Sigma_{A,\uparrow\uparrow}(t,t')
\end{pmatrix}.
\end{align}
This symmetry leads to a considerable reduction of the numerical complexity,
because one can make the $4\times4$ Dyson equation \eqref{latt dyson z}
$2\times2$ block-diagonal with the basis change
\begin{align}
\hat V
=
\frac{1}{\sqrt{2}}
\begin{pmatrix}
\hat 1 & \hat \sigma_x
\\
\hat \sigma_x & -\hat 1
\end{pmatrix}.
\end{align}
The symmetry \eqref{symmetry} implies
\begin{align}
\hat V 
\begin{pmatrix}
\hat Z_A & 0 
\\
 0 & \hat Z_B 
\end{pmatrix}
\hat V^\dagger
=
\begin{pmatrix}
\hat Z_A & 0 
\\
 0 & \hat Z_B 
\end{pmatrix},
\end{align}
and we have 
\begin{align}
\hat V 
\begin{pmatrix}
 0 & \hat \epsilon_\KK 
\\
 \hat \epsilon_\KK & 0
\end{pmatrix}
\hat V^\dagger
=
\begin{pmatrix}
\epsilon_\KK\hat\sigma_x & 0 
\\
 0 & -\epsilon_\KK\hat\sigma_x
\end{pmatrix}.
\end{align}
Thus the Dyson equation for the transformed $4\times4$ Green's functions
\begin{equation}
\tilde G_\KK(t,t') = \hat V \hat G_\KK(t,t') \hat V^\dagger.
\end{equation}
is block-diagonal: When we introduce the notation
\begin{align}
\tilde G_\KK 
\equiv 
\begin{pmatrix}
\hat G_\KK^{+}
&
0
\\
0 & \hat \sigma_x \hat G_\KK^{-} \hat \sigma_x 
\end{pmatrix},
\end{align}
(the $\hat \sigma_x$ in the second coefficients are introduced for convenience), 
the two blocks are obtained by solving two Dyson equations,
\begin{align}
\label{dyson gk+}
&\hat Z_A + \hat Z_A \conv  \epsilon_\KK \hat\sigma_x \conv \hat G_\KK^{+} = \hat G_\KK^{+}
\\
\label{dyson gk-}
&\hat Z_A - \hat Z_A \conv  \epsilon_\KK \hat\sigma_x \conv \hat G_\KK^{-} = \hat G_\KK^{-},
\end{align}
where we have again used the symmetry $\hat Z_B =\hat\sigma_x \hat Z_A\hat\sigma_x$ in the 
second equation. The back-transformation to the original basis, 
$\hat G_\KK=\hat V^\dagger \hat G_\KK \hat V$, gives
\begin{align}
\label{gk4 transf}
\hat G_\KK
=
\frac{1}{2}
\begin{pmatrix}
\hat G_\KK^{+} + \hat G_\KK^{-}  
&
(\hat G_\KK^{+}-\hat G_\KK^{-})\hat\sigma_x
\\
\hat\sigma_x( \hat G_\KK^{+}-\hat G_\KK^{-})
&
\hat\sigma_x( \hat G_\KK^{+} +\hat G_\KK^{-} )\hat\sigma_x
\end{pmatrix}.
\end{align}
The $\KK$-summed quantities \eqref{g-1}, \eqref{g1-1}, and \eqref{g2-1}  are thus obtained as
\begin{align}
\label{g-3}
\hat G_A 
&= 
\frac{1}{2L'}\sum_{\KK\in\bzC} \hat G_\KK^{+} + \hat G_\KK^{-}  
\\
\label{g1-3}
\hat G_A^{(1)} 
&= 
\frac{1}{2L'}\sum_{\KK\in\bzC} 
\epsilon_\KK \hat \sigma_x \conv  (\hat G_\KK^{+} - \hat G_\KK^{-} ) 
\\
\label{g2-3}
\hat G_A^{(2)} 
&= 
\frac{1}{2L'}\sum_{\KK\in\bzC} 
\epsilon_\KK \hat \sigma_x \conv  (\hat G_\KK^{+} + \hat G_\KK^{-} ) \conv \hat \sigma_x  \epsilon_\KK.
\end{align}
Here we have used Eqs.~\eqref{gab} and \eqref{gba} in Eqs.~\eqref{g1-1}, and \eqref{g2-1}, and then 
replaced $\hat G_{\KK\alpha,\alpha'}$ by the explicit expressions obtained from Eq.~\eqref{gk4 transf}.
Because all convolutions involve the time-local functions $\epsilon_\KK$, they are evaluated without
numerical cost.

The final set of DMFT equations, to be solved successively timestep after timestep, is thus given by:
(i) Solve {\em one} impurity model (the one on the $A$-lattice), i.e, compute $\hat G_A$ 
[Eq.~\eqref{gloc dmft}] from the action \eqref{dmft action} with hybridization $\hat \Delta_A$. 
(ii) Solve Eq.~\eqref{impdyson} for $\hat Z_A$. (iii) Solve two equations \eqref{dyson gk+}
and \eqref{dyson gk-} for $\hat G_\KK^+$ and $\hat G_\KK^-$. (iv) Evaluate the sums
Eq.~\eqref{g-3} to \eqref{g2-3}. 
(v) Compute the new hybridization function $\hat \Delta_A$
from Eq.~\eqref{get delta}.

Finally, the summation over $\KK$ is reduced to an integral over the density 
of states, as described in Ref.~\cite{Turkowski2005a}. We consider a cubic lattice with pure 
nearest neighbor hopping, and an electric field $\bm E(t)= E(t) (1,1,1,...) $ which is 
pointing along the body-diagonal of the unit cell. Then 
\begin{align}
\epsilon_\KK &= 
\frac{-2t^*}{\sqrt{2d}}
\sum_{\alpha=1}^d
\cos(k_\alpha-A(t))
\nonumber
\\
&=
\cos(A(t))
\epsilon^0_\KK
+
\sin(A(t))
\bar\epsilon^0_\KK,
\end{align}
where $\epsilon^0_\KK$ and $\bar \epsilon^0_\KK$ are 
band energies in the zero-field case, and 
$A(t)$ is the vector potential. The equations use a gauge with zero 
scalar potential, i.e., $E(t)=-\partial_t A(t)$, and the unit of the field
is $\text{hopping}/(e\times\text{lattice constant})$.
Since all functions depend on $\KK$ only via 
$\epsilon^0_\KK$ and $\bar \epsilon^0_\KK$, we can reduce the $\KK$ 
sum as
\begin{align}
\frac{1}{L'} \sum_{\KK\in\bzC}
f(\epsilon^0_\KK,\bar \epsilon^0_\KK)
=
\int 
\!
d\epsilon
\,d\bar \epsilon
\,
f(\epsilon,\bar\epsilon)
D'(\epsilon,\bar\epsilon),
\end{align}
with the density of states for the reduced zone
\begin{align}
D'(\epsilon,\bar\epsilon)
=
\frac{1}{L'} \sum_{\KK\in\bzC}
\delta(\epsilon-\epsilon^0_\KK)
\delta(\bar\epsilon-\bar \epsilon^0_\KK).
\end{align}
Because all points in the full BZ $\bz$ can be obtained by 
$\{ \KK,\KK+\QQ\}$ with $\KK\in\bzC$ and $\QQ=(\pi,\pi,...)$, 
and because $\epsilon^0_{\KK+\QQ}=-\epsilon^0_{\KK}$, we can choose 
the reduced BZ $\bzC$ as all $\KK$ with $\epsilon^0_{\KK}<0$. Hence we have
\begin{align}
D'(\epsilon,\bar\epsilon)
=
2
\Theta(-\epsilon)
D(\epsilon,\bar\epsilon),
\end{align}
where $D(\epsilon,\bar\epsilon)$ is the density of states for the full BZ. We will work in the limit of 
infinite dimensions, with $D(\epsilon,\bar\epsilon) = e^{-\epsilon^2}e^{-\bar\epsilon^2}$ \cite{Turkowski2005a}.

We close with the remark that the DMFT equations conserve the total spin along the direction of 
$\bm B$. The magnetic field $B_x$ thus determines only the time-independent expectation value 
of the initial field, while any time-dependence of a {\em homogeneous} magnetic field implies
a trivial time-dependent rotation of the Green's functions in spin space.

\section{Evaluation the exchange formulas}

In this section
we describe how we evaluate numerically the exchange interactions [Eq.~(2) of the main text] within DMFT.
In contrast to the evaluation of the DMFT self-consistency described above, this requires an explicit knowledge 
of the self energy. 
Below we discuss how the self energy $\mathit{\Sigma}_i(t,t')$ is evaluated using numerical derivatives. 
Once $\mathit{\Sigma}_i(t,t')$ is computed, the exchange interactions are evaluated by making the appropriate 
products and convolutions.

Within NCA, the self-consistent solution of the impurity model gives us direct access to the local Green function $G_{i}(t,t')$ 
and the hybridization function $\Delta_i(t,t')$ (where possible we omit the spin index $\sigma$). $\mathit{\Sigma}_i(t,t')$ is 
related to $G_{i}(t,t')$ and $\Delta_i(t,t')$ by the impurity Dyson equation:
\begin{eqnarray} 
\text{i}\partial_t G_i - \left[\Delta_i*G_i\right] &=& 1 + \left[\mathit{\Sigma}_i*G_i\right],
\end{eqnarray}
where $1$ indicates the delta function $\delta_\mathcal{C}(t,\tp)$ on the Keldysh contour. 
To be able to handle the equal-time discontinuities of the Green's functions and self energies analytically, 
we write the self-energy as $\mathit{\Sigma}_i=\bar{\mathit{\Sigma}_i}+\mathit{\Sigma}_i^\prime$. 
Here $\bar{\mathit{\Sigma_i}}(t,t')=\delta_\mathcal{C}(t,\tp)\mathit{\Sigma}^H_i(t)$ is the Hartree 
component of the self-energy and $\mathit{\Sigma}_i^\prime(t,\tp)$ is the part of the self-energy 
which is finite at $t=t'$.

The Hartree component is computed by invoking a first numerical derivative, indicated as $\partial_{\mathrm{N}t}$, which is evaluated only for $t=\tp\pm\varepsilon$. Denoting $F_i=\mathit{\Sigma}_i*G_i$ we write the Dyson equation as:
\begin{eqnarray}
F_i &=& \text{i}\partial_tG_i - 1 - \Delta_i*G_i \label{e:f}\\
&=&(\text{i}\partial_{\mathrm{N}t}G_i+1) - 1 - \Delta_i*G_i  = \text{i}\partial_{\mathrm{N}t}G_i - \Delta_i*G_i.\nonumber
\end{eqnarray}
$\mathit{\Sigma}_i^{H\bar{\sigma}}(t)$ now follows directly from the equal time contribution $F_i(t,t)$. On the real-time axis we have
\begin{equation}
F^{>\sigma}_i(t,t)-F^{<\sigma}_i(t,t)=-\text{i}\,\mathit{\Sigma}_i^{H\bar{\sigma}}(t),\label{e:sigmah}
\end{equation}
where we used that $G^>(t,t)-G^<(t,t)=-\text{i}$. On the Matsubara axis we have equivalently $F^{\sigma}_i(0)+F^{\sigma}_i(\beta)=-\mathit{\Sigma}_i^{H\sigma}(0)$. 


To compute the component $\mathit{\Sigma}_i^\prime(t,\tp)$ we invoke a second derivative and write the local $T$ matrix as:
\begin{eqnarray}
T_i &=& \text{i}\partial_tF^\dagger_i-\Delta_i*F^\dagger_i = \left[\text{i}\partial_tG_i-\Delta_i*G_i\right]*\mathit{\Sigma}_i \nonumber \\
&=& \left[\text{i}\partial_{\text{N}t}G_i+1-\Delta_i*G_i\right]*\mathit{\Sigma}_i \label{e:ti1} \\
&=& \text{i}\partial_{\text{N}t}F^\dagger_i-\Delta_i*F^\dagger_i + \mathit{\Sigma}_i. \nonumber
\end{eqnarray}
In addition, we have
\begin{eqnarray}
T_i &=& \left[\text{i}\partial_tG_i-\Delta_i*G_i\right]*\mathit{\Sigma}_i = \mathit{\Sigma}_i+\mathit{\Sigma}_i*G_i*\mathit{\Sigma}_i  \nonumber \\
&=& \bar{\mathit{\Sigma}}_i+\mathit{\Sigma}_i^\prime + F_i*(\bar{\mathit{\Sigma}}_i+\mathit{\Sigma}_i^\prime) \label{e:ti2} \\
&=& \bar{\mathit{\Sigma}}_i+F_i\,\mathit{\Sigma}^H_i + (1 + F_i)*\mathit{\Sigma}_i^\prime. \nonumber
\end{eqnarray}
By subtracting Eq.~\eqref{e:ti1} and Eq.~\eqref{e:ti2} we obtain:
\begin{equation}
(1 + F_i)*\mathit{\Sigma}_i^\prime = \text{i}\partial_{\text{N}t}F^\dagger_i-\Delta_i*F^\dagger_i - F_i\,\mathit{\Sigma}^H_i. \label{e:sigmaprime}
\end{equation}
Finally, the self-energy $\mathit{\Sigma}_i^\prime$ is evaluated by inverting Eq.~\eqref{e:sigmaprime}.
This integral equation corresponds to a Volterra equation of the second kind, which is numerically 
well conditioned.

\bibliography{rmpbib}

\end{document}